\documentstyle[axodraw,colordvi]{frank}

\newcommand{\mrm}[1]{\mbox{\rm #1}}
\newcommand{\beq}{\begin{equation}}
\newcommand{\eeq}{\end{equation}}

\newcommand{\bea}{\begin{eqnarray}}
\newcommand{\eea}{\end{eqnarray}}

\newcommand{\rfn}[1]{(\ref{#1})}
\newcommand{\Eq}[1]{Eq.~(\ref{#1})}

\newcommand{\np}[1]{Nucl. Phys. {\bf #1}}
\newcommand{\pl}[1]{Phys. Lett. {\bf #1}}
\newcommand{\pr}[1]{Phys. Rev. {\bf #1}}
\newcommand{\prl}[1]{Phys. Rev. Lett. {\bf #1}}
\newcommand{\zp}[1]{Z. Phys. {\bf #1}}

\newcommand{\ptp}[1]{Prog. Theor. Phys. {\bf #1}}

\newcommand{\sm}{standard model}
\newcommand{\cm}{centre of mass}
\newcommand{\xs}{cross section}

\newcommand{\lc}{linear collider}
\newcommand{\pe}{\mbox{$e^+e^-$}}
\newcommand{\ee}{\mbox{$e^-e^-$}}
\newcommand{\mm}{\mbox{$\mu^-\mu^-$}}

\newcommand{\bil}{bilepton}

\newcommand{\guts}{grand unified theories}

\def\lsim{\mathrel{\vcenter{\hbox{$<$}\nointerlineskip\hbox{$\sim$}}}}
\def\gsim{\mathrel{\vcenter{\hbox{$>$}\nointerlineskip\hbox{$\sim$}}}}

\begin{document}

\begin{flushright}
FTUV/97-34\\
IFIC/97-33\\
June 1997
\end{flushright}

\vfill

\begin{frontmatter}
\title{Lower Bounds on Bilepton Processes  \\
     at \ee\ and \mm\  Colliders}
\author{Martti Raidal}
\address{{\tt raidal@titan.ific.uv.es}\\
        Department of Theoretical Physics and IFIC, \\
        University of Valencia,
        E-46100 Burjassot, Valencia, Spain}
\begin{abstract}
We show that the \xs\ of at least one of the s-channel processes \linebreak
$\ee (\mm)\rightarrow \ell_i^-\ell_i^-,$ $i=e,$ $\mu,$ $\tau,$
mediated by a doubly-charged scalar
triplet bilepton is bounded from below and 
observable at a linear or muon collider if one of the light
neutrinos has a mass in the range where it is required to be unstable 
by  cosmological considerations. The result is model independent.
We therefore stress the importance of the \ee\ and \mm\ collision modes 
of the future colliders for discovering new physics.
\end{abstract}


\end{frontmatter}

\vfill
\clearpage

The conservation of lepton flavour which has been tested with 
great accuracy at low energies  is an important feature
of the \sm.  However, at  energies of TeV range  to be offered
by electron linear  \cite{lc} and muon colliders \cite{mucol} 
lepton flavour may turn out to be not exact symmetry. 
The \ee\ and \mm\ running modes~\cite{e-e-} of the colliders 
are particularly suitable for discovering this type of new physics 
since the initial states carry double electron  and muon 
 number, respectively. 

One of the many promising processes 
which can be studied in these modes 
is the s-channel production of lepton pairs mediated 
by doubly-charged \bil s.
These particles are predicted 
by wide range of  extensions of the \sm,
such as \guts~\cite{guts},
theories with enlarged Higgs sectors~\cite{higgs},
theories which generate neutrino Majorana masses~\cite{maj}
as well as technicolour theories~\cite{tc} and 
theories of compositeness~\cite{comp}.
Particularly interesting among them are the theories containing scalar
triplet \bil s since they provide a framework for the understanding 
of the smallness of the masses of the ordinary neutrinos via
the see-saw mechanism \cite{seesaw}.

In this letter we consider the processes 
$\ee (\mm)\rightarrow \ell_i^-\ell_i^-,$ $i=e,$ $\mu,$ $\tau,$
\cite{ee2bil,jack} induced by a scalar triplet bilepton 
in a model independent way. We show that there are lower 
bounds on the \xs\ of the processes  and at least one of
them can be detected in the planned  lepton colliders provided 
that one of the light neutrinos, either  $\nu_\mu$ or $\nu_\tau,$
has a mass in the range  for which the constraint from the 
energy density of the present Universe requires it to be unstable.

We define \bil s to be bosons 
which couple to two leptons
and which carry two units of lepton number.
Their interactions need not necessarily conserve lepton flavour,
but otherwise we demand the symmetries of the \sm\ to be respected.
The most general $SU(2)_L\times U(1)_Y$ invariant 
renormalizable dimension four Lagrangian of this kind for triplet 
scalar bileptons 
$B_{3}^{0},\ B_{3}^{+}$ and $B_{3}^{++}$ 
is given by~\cite{cd}
\bea
\label{lag}
{\cal L}
&=& \lambda_3^{{ij}}\left(
 B_{3}^{0} \, \bar\nu^c_i P_L \nu_j 
- \frac{B_{3}^{+}}{\sqrt{2}}  \left( \bar \ell^c_i P_L \nu_j + 
\bar \ell^c_j P_L \nu_i \right)
-  B_{3}^{++} \bar \ell^c_i P_L \ell_j \right)
+ \mbox{ h.c.}~,
\eea
where the indices $i,j=e,\mu,\tau$ stand for the lepton flavours and
the chirality projection operators
are defined as $P_{R,L} = (1\pm\gamma_5)/2$.
In the following  we shall drop the subscripts 3 denoting 
the dimension of the bilepton representation everywhere.

Due to large bilepton masses $m_B$  the present low energy 
 experiments can only constrain their effective  couplings
of a generic form  
$G=\lambda^2/m^2_B.$  Negative results in searches for 
 the lepton flavour violating processes 
$\ell_l\rightarrow 3\ell_f$ and $\ell_l\rightarrow \gamma \ell_f,$
where $l=\mu, \; \tau,$ and $f=e,\; \mu,$ put orders of magnitude
more stringent bounds on  off-diagonal bilepton couplings than 
one obtains from M{\o}ller scattering and $(g-2)_\mu$ studies as well as
from the searches for muonium-antimuonium conversion for diagonal
couplings $\lambda^{ee}$ and  $\lambda^{\mu\mu}$ 
 \cite{cd,willy,lnv}. To date there is no constraints on  
$\lambda^{\tau\tau}$ without involving off-diagonal elements. 
Following the present phenomenology 
 we can approximate the bilepton coupling matrix
$\lambda^{ij}$ to be diagonal (small off-diagonal elements will
not change our conclusions). Due to our ability to rotate 
lepton fields by a  phase we can 
choose the couplings $\lambda^{ii}$ in \Eq{lag}
to be real without loss of generality.

While the linear and muon colliders will probe the couplings and 
masses of the bileptons orders of magnitude more tightly than any of the 
present experiments \cite{jack,lam} it is possible that due to  
small $\lambda$'s or high bilepton  masses no positive signal 
will be detected.
However, this may not be the case if neutrinos are massive as
predicted by most of the extensions of the \sm. 

The masses 
and lifetimes of the neutrinos are constrained by the requirement that 
the energy density of them in the present Universe does not exceed the 
upper limit on the total energy density of the Universe \cite{uus}. If the
sum of light  neutrino masses exceeds $\sim 90$ eV at least one of 
them has to be unstable. This may, indeed, be the case for $\mu$-
or $\tau$-neutrinos since 
the present upper limits on neutrino masses are $m_{\nu_e}\lsim 10$ eV,
$m_{\nu_\mu}\lsim 170$ keV \cite{pdb} and $m_{\nu_\tau}\lsim 18$ MeV 
\cite{aleph}.
The lifetime of such an unstable neutrino $\nu_l$
must satisfy the requirement
\cite{eight} 
\bea
\tau_{\nu_l} \lsim 8.2 \cdot 10^{31} \; \mbox{MeV}^{-1} \left(
\frac{100 \; \mbox{keV}}{m_{\nu_l}} \right)^2 .
\label{constr}
\eea
The  $\nu_l$'s  can decay either radiatively 
$\nu_l \to  \nu_f \gamma, \; \nu_f\gamma\gamma $ 
or at tree level $\nu_l \to 3\nu_f$ via neutral
bilepton $B^0$ or $Z'$ exchange. 
The radiative decay modes are highly suppressed \cite{rad}
and cannot satisfy  the constraint (\ref{constr}). The same is also 
true for $Z'$ contribution to $\nu_l \to 3 \nu_f$ decay \cite{zprim}.
Therefore, we are left with the decays
$ \nu_l   \to   3 \nu_f $ induced by the 
$B^0$ exchange. The effective Hamiltonian for this process
is given by
\bea
H = \frac{ G_{0}^{lf}}{\sqrt{2}} \overline{\nu}_f \gamma^\mu 
\left( 1 - \gamma_5 \right) \nu_f \, \overline{\nu}_f  \gamma_\mu
 \left( 1 - \gamma_5 \right) \nu_l  +  \mbox{h.c.}\;, 
\label{g0}
\eea
where 
$G_{0}^{lf} = \sqrt{2} \lambda_{ff} 
(\lambda_{ff} K_{fl}+\lambda_{ll} K_{lf})
/(4 m^2_{B^0}). $ 
Note that the neutrino decays only due to the neutrino mixings
presented by a mixing  matrix $K_{ij}$ and
the processes $\ell_l \to 3 \ell_f $ are forbidden as required by the 
current phenomenology.

Just from the   $\nu_l$ lifetime, 
$\tau_{\nu_l}^{-1} = 2 G_{0}^2 m_{\nu_l}^5/(192  \pi^3),$
and the constraint \rfn{constr} we obtain a {\it lower}  bound
on the effective coupling $G_0$ as
\beq
G_0^{lf}\gsim 1.9\cdot 10^{3}\; \mrm{TeV}^{-2}\;
\left( {\mrm{keV}\over m_{\nu_l}} \right)^{\frac{3}{2}}.
\label{g0bound}
\eeq
Numerically the minimum values of $G_0$ depend on whether the
unstable neutrino is of $\mu$ of $\tau$ type (we assume that 
only one neutrino is unstable) and are given by 
$G_0^{\mu e}\gsim 8.6\cdot 10^{-1}$ TeV$^{-2}$ and 
$G_0^{\tau f}\gsim 8.0 \cdot 10^{-4}$ TeV$^{-2}.$

\begin{figure}[htb]
\unitlength.5mm
\SetScale{1.418}
\begin{boldmath}
\begin{center}
\begin{picture}(70,40)(0,0)
\ArrowLine(0,0)(15,15)
\ArrowLine(0,30)(15,15)
\DashLine(15,15)(45,15){1}
\ArrowLine(45,15)(60,30)
\ArrowLine(45,15)(60,0)
\Text(-2,0)[r]{$\ell_f^-$}
\Text(-2,30)[r]{$\ell_f^-$}
\Text(62,0)[l]{$\ell_i^-$}
\Text(62,30)[l]{$\ell_i^-$}
\Text(30,23)[c]{$B^{--}$}
\end{picture}
\end{center}
\end{boldmath}
\bigskip
\caption{
  Lowest order Feynman diagram 
  inducing the processes $\ell^-_f \ell^-_f\to \ell_i^- \ell_i^-$.
  Index $i$ goes over all flavours while $f$ denotes only 
  $e$ or $\mu.$
}
\bigskip
\label{fee2mm}
\end{figure}
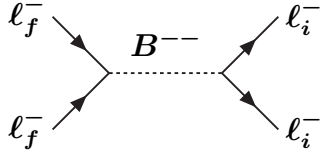

Let us now turn to studies of collider physics.
Leptons can be pair-produced in \ee\ and \mm\ collisions
via the $s$-channel exchange of a doubly-charged  \bil,
as depicted in Fig.~\ref{fee2mm}.
\bigskip
Assuming fully left polarized  incoming beams
(in reality polarization rates exceeding 90\% are achievable \cite{zdr})
the total \xs s of the processes are given by
\beq
\label{totxs}
\sigma_{fi}~
=~
{\lambda_{ff}^2\lambda_{ii}^2 \over 2\pi}~
{s \over (s-m_{B^{++}}^2)^2 + m_{B^{++}}^2\Gamma_{B^{++}}^2}~,
\eeq
where the bilepton leptonic width is 
$\Gamma_{B^{++}}=
\sum_{i}\lambda_{ii}^2  m_{B^{++}}/8\pi$
\cite{ee2bil,bildec}.
If the doubly-charged members of the multiplets
turn out to be heavier 
than the singly-charged ones,
the non-leptonic decay mode
$B^{--} \to B^-W^-$
can possibly also contribute to the total width.
However,
only very heavy \bil s
can realistically accommodate a mass splitting 
exceeding the mass of the $W$ boson. Also the decay mode
$B^{--}\to W^-W^-$ may contribute to the bilepton width 
but it is strongly suppressed by the neutral bilepton 
vev $v_{B^0},$
which  is constrained to be small due to its contribution to the $\rho$
parameter \cite{gunion}.
With the current  constraints
on the possible values of the decay width 
the $B^{--}$ resonance peak at the  colliders is very prominent 
and the resonant \xs\ is much larger than the off-resonance one \cite{jack}.
Therefore we can study only the most conservative situation where
the processes $\ell_f^-\ell^-_f\rightarrow \ell^-_i\ell^-_i$ 
are mediated by far off-resonance virtual bilepton.
For completness we shall later comment also on the production of $W^-W^-$
in these collisions.

If the doubly charged bilepton is very heavy, $s\ll m_{B^{++}}^2,$ 
its interaction with leptons can be characterized by an
effective Hamiltonian
\bea
H = \frac{ G_{++}^{fi}}{\sqrt{2}} 
\overline{\ell}_i \gamma^\mu 
\left( 1 - \gamma_5 \right) \ell_f \, \overline{\ell}_i  \gamma_\mu
 \left( 1 - \gamma_5 \right) \ell_f  +  \mbox{h.c.}\;,
\eea
where 
$G_{++}^{fi} = \sqrt{2}\lambda_{ff}\lambda_{ii}/(8 m^2_{B^{++}}). $
In this formalism the \xs\ \rfn{totxs} takes a form
\beq
\label{totxs2}
\sigma_{fi}~
=~
{16\; (G_{++}^{fi})^2 \over \pi}~s ~.
\eeq
The effective couplings $G_{++}$ are related to $G_0$ as
\beq
G_{++}^{fi}=\frac{G_0^{lf}}{2} 
\frac{m^2_{B^0}}{m^2_{B^{++}}}
{\lambda_{ii}\over \left( 
\lambda_{ff} K_{fl}+\lambda_{ll} K_{lf} \right)} \; .
\label{grelation}
\eeq 
Clearly, since some of $G_0$'s are bounded from below
also some of $G_{++}$'s cannot be arbitrarily small
leading to non-vanishing processes at colliders. 
For large neutrino mass differences the present limits on the
neutrino mixings are 
$| K_{e\mu} |=| K_{\mu e} |\lsim 2.8\cdot 10^{-2},$
$| K_{\mu\tau} |=| K_{\tau \mu} |\lsim 3\cdot 10^{-2}$
and
$| K_{e\tau} |=| K_{\tau e} |\lsim 0.2$ \cite{pdb,naus}.
Also the mass splitting between $B^{++}$ and $B^0,$ which belong to
the same $SU(2)_L$ multiplet,  is strongly bounded 
from the experimental value of the  parameter
$\rho = 1 + \rho_\theta + \rho_B, $ 
where $\rho_\theta$ is a
correction  due to the mixing of $Z^0$ with 
a new neutral gauge boson
(which we are neglecting here) and $\rho_B$ comes from
the bilepton contribution to the $Z^0$ and $W^{\pm}$ mass. It is
given by \cite{gunion}
\bea
\rho_B= \frac{G_F}{4 \sqrt{2} \pi^2} \left[ f_{(B^0,
B^+)} + 
f_{(B^+,B^{++})} 
\right] \equiv \frac{3 G_F}{8 \sqrt{2} \pi^2}
\Delta m ^2 , 
\eea
where
$ f_{(x,y)}=m_x^2 +m_y^2 - 2 m_x^2 m_y^2 \ln (
m_y^2/m_x^2 )/(m_y^2-m_x^2). $
Studies of the new contributions to the $\rho$ parameter 
have provided the upper bounds  
$\Delta m^2  \leq (76 \; \mbox{GeV} )^2, $ 
$(98 \; \mbox{GeV} )^2,$ $ (122 \; \mbox{GeV} )^2$ \cite{pdb} 
for the \sm\ Higgs masses $m_H = 60,$ 300 and 1000 GeV, respectively, at  
90$\%$ C.L. Therefore, for the interesting range of bilepton masses 
of  1 TeV and higher 
the  ratio   $m_{B^0}/m_{B^{++}}$ cannot differ from unity more than
$\sim 10$-20\% even for $m_H=1$ TeV.

Let us now study the implications of the bound \rfn{g0bound}
on $B^{--}$ processes at the colliders. 
To estimate the  discovery potential,
we use the scaling relation $  {\cal L}_{e^-e^-}=3.25\cdot 10^7~s~$ 
for the \ee\ and \mm\ luminosities
which closely corresponds 
to a luminosity of 25 fb$^{-1}$ at $\sqrt{s}=0.5$ TeV
and scales like the square of the \cm\ energy.
This choice for the luminosity
is dictated by the latest \pe\ \lc\ design report~\cite{zdr}
and the fact that the \ee\ mode 
will approximately suffer a 50\%\ luminosity reduction
because of the anti-pinch effect \cite{jim}.

Concerning  the background
 we have to deal with two different types of processes. 
While the lepton number violating processes have no background
from the \sm\ then \ee\ and \mm\ elastic scatterings take place
also without bileptons. However, due to the interference
with the \sm\ graphs the bilepton effects are enhanced in the latter case 
which compensates the existence of the background.   
If we assume that observing one flavour violating  event
already constitutes a discovery,
we need an average number of 
$-\ln(1-p)$
Poisson distributed events 
such that {\em at least} 1 event
is observed with probability $p$.
Hence, a predicted average of at least 3 events
is needed to guarantee a discovery with 95\%\ confidence.
In this case the 
minimal testable $G_{++}$'s following from the \xs\ \rfn{totxs2}
are
\beq
G_{++}^{fl}(\mrm{min})=\frac{1.4\cdot 10^{-4}}{s} 
\;\mrm{TeV}^{-2}\;,
\label{gpmin2}
\eeq  
where $f\neq l$ and $s$ is expressed in  TeV$^2$.
In the case a tau lepton is produced we assume its reconstruction
efficiency to be 65\% and this value should be divided by 0.65.
The situation $f=l=e,\;\mu$ has been studied in \cite{cd}
where the Cramer-Rao limit,
$\chi^2_\infty={\cal L}\int dt 
[d\sigma(\lambda)/dt-d\sigma(\lambda=0)/dt]^2/[d\sigma(\lambda=0)/dt],$
has been computed. At  95\% confidence level $\chi^2_\infty=3.84,$ and 
one obtains
\beq
G_{++}^{ff}(\mrm{min})=\frac{8\cdot 10^{-5}}{s} 
\;\mrm{TeV}^{-2}\;.
\label{gpminee}
\eeq

Independently of which neutrino has the large mass and  decays to three
lighter neutrinos we can always choose to study 
the process $\ell_f^-\ell_f^-\to\ell_l^-\ell_l^-$ at the 
linear or muon collider
and to constrain the relevant couplings $\lambda_{ff},$
$\lambda_{ll},$  $l\neq f,$ $i=l$ in \Eq{grelation}.
For the numerical estimates  we choose the case if 
$\nu_\tau\to3\nu_e$
 since the  present experimental constraints on 
$|K_{\tau e}|$ and $m_{\nu_\tau}$
give us the most conservative limits. 
Using the numerical quantities determined above we obtain from 
Eqs \rfn{g0bound}, \rfn{grelation} 
\beq
G_{++}^{e\tau}\gsim 2\cdot 10^{-3} \frac{\lambda_{\tau\tau}}
{\lambda_{ee}+\lambda_{\tau\tau}} 
\;\mrm{TeV}^{-2}\;,
\label{gpminte}
\eeq  
which in comparison with \Eq{gpmin2} 
implies that the process $e^-e^-\to \tau^-\tau^-$ should be
detected at the 1 TeV \lc\ unless 
$\lambda_{\tau\tau}/\lambda_{ee}\lsim  10^{-1}.$
On the other hand, if this is the case then
\beq
G_{++}^{ee}\gsim 2\cdot 10^{-3}  
\;\mrm{TeV}^{-2}\;,
\label{gpminee2}
\eeq  
and \Eq{gpminee} suggests that the  excess of the electron 
pairs due to the s-channel bilepton production will be detected. 
Note that the positive signal should be seen if $\sqrt{s}\gsim 0.3$ TeV
which is below the planned initial energy of the \lc .

Similarly, if 
$\nu_\tau\to 3\nu_\mu$ then  $\mu^-\mu^-\to \tau^-\tau^-$
should be seen unless $\lambda_{\tau\tau}/\lambda_{\mu\mu}\lsim 10^{-3}$
($\sqrt{s}_{\mu\mu}=4$ TeV used), and if 
$\nu_\mu\to 3\nu_e$ then  $e^-e^-\to \mu^-\mu^-$
should be seen unless $\lambda_{\mu\mu}/\lambda_{ee}\lsim 10^{-5}.$
 Suppression of the flavour violating processes by   
small $\lambda_{ll}/\lambda_{ff}$ would mean that
the \xs s of the flavour conserving processes 
$\ell_f^-\ell_f^-\to \ell_f^-\ell_f^-$ 
exceed the minimal observable limit \rfn{gpminee} by  orders of
magnitude.
Therefore, the bilepton mediated s-channel processes cannot be missed 
at future facilities.

One should also note that at the time colliders start to operate 
new experimental data on neutrino mixings and masses will be available. 
The largest improvements in experimental sensitivity 
can be expected in tau neutrino physics.
Proposed E803 and NAUSICAA  experiments at Fermilab will respectively have
about one and two orders of
magnitude higher sensitivity
to $|K_{e\tau}|$ and $|K_{\mu\tau}|$  than  the present limits \cite{naus}.
Should these experiments give negative results then,
together with improvements of $m_{\nu_\tau}$ determination 
in tau factories,
the bounds on $G_{++}$'s will rise about a factor of hundred.

Finally, let us comment on the possibility of observing 
the processes $\ell_f^-\ell_f^-\to W^-W^-$ \cite{ww}. 
The $B^{++}W^-W^-$
vertex is proportional to the vev $v_{B^0}$ which is experimentally
constrained to be below a few GeV \cite{gunion}. 
However, there is no fundamental
reason that it is exactly zero. Therefore, for  the energies where
$W$ pair production is not kinematically suppressed, 
$M^2_W\ll s\ll M^2_B,$  one gets
\beq
\sigma (\ell_f^-\ell_f^-\to W^-W^-) \sim 
\sigma (\ell_f^-\ell_f^-\to \ell_i^-\ell_i^- ) 
\frac{g^4 v^2_{B^0} s}{\lambda_{ii}^2 M^4_W}\;. 
\eeq
In the case of  small $\lambda_{ii}$
the \xs\ of $\ell_f^-\ell_f^-\to W^-W^-$ may be enhanced by a 
factor of $g^4 v^2_{B^0}s/(\lambda_{ii}^2 M^4_W)$
and provide an observable amount of lepton number violating
events.

In conclusion, we have shown that if one neutrino has 
a mass exceeding about $\sim 90$ eV there is such a lower bound on
the \xs\ of {\it at least} one of the processes  
$\ell_f^-\ell_f^-\to \ell_i^-\ell_i^-,$ mediated by the doubly charged 
scalar triplet bilepton $B^{--},$ that the process is observable at 
future \ee\ or \mm\ colliders.

We stress that this result is very general. We have not used any
model dependent relation for the bilepton couplings nor for
the neutrino masses. Our conclusion is also independent of the 
exact values of bilepton masses since only the effective
couplings $G^{fi}$ are constrained by the 
analyses. For our considerations it is important that triplet bileptons
and massive neutrinos do exist. Therefore it appears to be  difficult to 
avoid the lower bounds  if the collider 
parameters will be close to the presently designed ones.
Small changes in the  collider parameters and cosmological bounds,
small non-zero off-diagonal bilepton couplings as
well as accidental cancellations between the used parameters
may change our numerical values by a factor of 2-3, or so, but not by orders
of magnitude what is required to avoid our conclusions.

We thank F. Cuypers and P. Zerwas for valuable comments on the manuscript.
This work is supported by the CICYT under grant AEN-96-1718.

\end{document}